\documentclass[twocolumn]{aastex62}
\usepackage{amsmath}

\graphicspath{{figures/}}

\received{}
\revised{}
\accepted{}

\shorttitle{Disk Polarization with Big Grains}
\shortauthors{Yang and Li}

\begin{document}

\title{Scattering-Induced Disk Polarization By Millimeter-Sized Grains}

\correspondingauthor{Haifeng Yang}
\email{yanghaifeng@tsinghua.edu.cn}

\author[0000-0002-8537-6669]{Haifeng Yang}
\altaffiliation{C.N. Yang Junior Fellow}
\affil{Institute for Advanced Study, Tsinghua University, Beijing, 100084, China}

\author{Zhi-Yun Li}
\affil{Department of Astronomy, University of Virginia, Charlottesville, VA 22904, USA}



\begin{abstract}
Spatially resolved (sub)millimeter polarization has been detected by ALMA in an increasing number of disks around young stellar objects. The majority of the observations show polarization patterns that are consistent with that expected from scattering by dust grains, especially at the shortest wavelength band for ALMA polarization ($\lambda \approx 870\,\mathrm{\mu m}$). The inferred sizes of the grains responsible for the scattering-induced polarization are typically of order $100\,\mathrm{\mu m}$, which is very different from the millimeter size commonly inferred from the dust opacity index $\beta$. In an effort to resolve this discrepancy, 
we first introduced the so-called ``Coplanar Isotropic Radiation Field" approximation, which enables the computation of the (signed) polarization fraction (negative means polarization reversal) analytically. 
With an oft-adopted dust composition (used, e.g., by DSHARP), we find that models with big dust grains produce very small polarization
with reversed orientation (relative to the well-known Rayleigh scattering by small particles), which hasn't been observed. 
The semi-analytic results are validated through Monte Carlo radiative transfer simulations. In these models, 
the ``correct" (not reversed) polarization orientation and the small $\beta$ index are mutually exclusive. 
To resolve this tension, we explore a wide range of dust models, parameterized by their complex refractive index $m=n+ik$. 
We find that both the fraction of the polarization and whether it is reversed or not depend on the refractive index in a complex way, and this dependence is mapped out on an $n-k$ plane for a representative dust size distribution (MRN with $a_\mathrm{max}=3$~mm) and wavelength of $870\,\mathrm{\mu m}$.  In particular, 3~mm-sized grains made of refractory organics produce polarization that is reversed, whereas grains of the same size but made of absorptive carbonaceous materials can produce a percent-level polarization that is not reversed; the latter may alleviate the tension between the grain sizes inferred from the scattering-induced polarization and the opacity index $\beta$. We conclude that scattering-induced polarization has the potential to probe not only the sizes of the grains but also their composition. 
\end{abstract}



\section{Introduction} \label{sec:intro}
Magnetic fields play a crucial role in the evolution and dynamics of an accreting
protoplanetary disk, through either magneto-rotational instability \citep{Balbus1991} or 
magnetically-driven wind \citep{Blandford1982}. Direct evidence for the magnetic field in the 
disk, however, has been lacking. 

Polarized light in the (sub)millimeter wavelengths has been established as a promising tool in 
probing celestial magnetic fields on larger scales ($\gtrsim 1000\,\mathrm{AU}$), from molecular
clouds \citep{Planck2015XIX,Fissel2016}, to star forming cores (see, e.g., reviews by \citealt{Pattle2019} and \citealt{Hull2019}). However, when it comes to the disk scale
($\sim 100\,\mathrm{AU}$ or less), the picture becomes more complicated. The very first resolved
polarization map of a classical T Tauri disk using the Combined Array for Research in Millimeterwave
Astronomy (CARMA), 
reveals a uniform polarization pattern which would imply an unphysical magnetic field configuration
in a differentially rotating system \citep{Stephens2014}. Since then, and thanks to the Atacama Large Millimeter Array (ALMA), 
many disk polarization maps have been made, leading to rapid observational progress in the field. 
The origins of 
these disk polarizations, however, remain unclear. Alongside the classical
magnetic alignment interpretation, there exist at least three alternatives, scattering-induced
polarization, radiative alignment, and mechanical alignment. 

Although each one of the four aforementioned mechanisms has some observational support in certain systems,
there is not a single mechanism that can explain all observed polarization in protoplanetary
disks. 
Alignment with respect to the local radiation anisotropy (``k-RAT alignment'' thereafter) is
best supported by the azimuthal polarization pattern observed at ALMA Band 3 in the HL Tau system
\citep{Kataoka2017}. However, it predicts a strong azimuthal variation
of polarization and circular pattern (rather than elliptical pattern)
\citep{Yang2019}. 
There is some tentative evidences for 
alignment with respect to the magnetic field, through either 
Radiative Alignment Torques (``B-RAT alignment''; \citealt{Lazarian2007}), or recently proposed
Mechanical Alignment Torques \citep{Hoang2018}, in, e.g.,  the IRAS 4A 
system at cm wavelengths \citep{Cox2015,Yang2016b} and BHB 07-11 (F. Alves et al. 2018) at (sub)millimeter wavelengths. 
But there is no well resolved system that matches the 
theoretical expectations (see, e.g., \citealt{Cho2007,Yang2016b,Bertrang2017}) assuming the widely expected disk 
toroidal magnetic 
field yet \citep{Flock2015}.
Mechanical alignment has recently received some attention. \cite{Hoang2018} claims that under MATs,
grains can be aligned with respect to local dust-gas streaming direction, in the
case of a weak or zero magnetic field, even if the velocity difference is sub-sonic. Within this picture,
\cite{Kataoka2019} investigated the direction of streaming velocities for dust grains with different
Stokes numbers, and the resulting polarization orientations. They found that their
polarization pattern in the order-of-unity Stokes number case resembles that observed by
\cite{Alves2018} in BHB07-11.
The BHB07-11, however, is a binary system, and we expect more complicated velocity fields than
the simple one assumed in \cite{Kataoka2019}. \cite{Yang2019} investigated the observational features of another mechanical 
alignment mechanism, the Gold mechanism \citep{Gold1952}, to address the circular versus elliptical pattern problem in the ALMA Band 3 polarization observations of HL Tau disk.
However, they failed to explain the non-existence of strong azimuthal variation, and suggested 
the scattering by dust grains aligned under the Gold mechanism may be the origin of the 
polarization at ALMA Band 3 in the HL Tau system.

The scattering-induced polarization is quite different
from alignment-based mechanisms discussed above. It is the mechanism that has the strongest 
observational support so far. Soon after it was
initially proposed by \cite{Kataoka2015}, \cite{Yang2016a} first pointed out that 
self-scattering in the Rayleigh regime will produce uniform polarization patterns in an 
inclined disk system (see also \citealt{Kataoka2016a}, and discussions in Sec.~\ref{subsec:cirf}). 
This uniform polarization pattern has been observed in many systems, with HL Tau 
\citep{Stephens2017}, and IM Lup \citep{Hull2018}, and HD 163296 \citep{Dent2019} 
as the best and well-resolved examples,
plus many others, e.g. DG Tau \citep{Bacciotti2018}, HH80/81 \citep{Girart2018}, 
VLA1623 \citep{Harris2018}, HH 111 and HH 212 \citep{Lee2018}, and RY Tau \citep{Harrison2019}. 
The predicted (and observed) uniform polarization pattern is almost impossible to produce
for any alignment-based mechanism. These alignment-based mechanisms rely on the dichroic 
emission of aspherical dust grains aligned with respect to some sort of local field, 
e.g. the magnetic field, which is usually varying its direction in a rotating protoplanetary 
disk.
At this time, the observed uniform polarization pattern is a feature unique to the  scattering-induced polarization, and the wide-spread detection of such a pattern is a strong evidence for the mechanism. 

Despite the success of the scattering-induced polarization, it is in serious tension with previous
work using the ``$\beta$ index" to probe grain sizes (see PPVI review
by \citealt{Testi2014} and references therein).  
Thanks to the strong dependence of scattering opacity on grain sizes, one can probe the grain
sizes in protoplanetary disks. In order to reproduce the observed polarization fraction at
ALMA Band 7 ($\sim 870\rm\,\mu m$), the optimal grain size is about $\sim 140\rm\,\mu m$ when
the size parameter ($x\equiv 2\pi a/\lambda$) is on order of unity. 
Previous work using this method all found grain sizes not too far from this
value ($\sim 100\rm\, \mu m$; \citealt{Yang2016a}, \citealt{Kataoka2016a}, \citealt{Hull2018},
\citealt{Dent2019}). 

On the other hand, the well-established ``$\beta$ index'' method can probe the grain size
in protoplanetary disk independently. If the inferred opacity index $\beta$ is smaller than
1 in a system at a certain wavelength $\lambda$, the emission at that wavelength should be dominated by grains of at least $3\lambda$ in size \citep{Draine2006,Testi2014}. Small opacity index $\beta$ at 
millimeter wavelengths can then serve as a probe for big dust grains with sizes of millimeters 
or even centimeters (e.g., \citealt{Perez2012,Perez2015}). 
For example, \cite{ALMA2015} reported resolved $\beta$ index in HL Tau disk with values mostly smaller than one between ALMA Band 7 ($\sim 0.87 \rm\, mm$) and Band 6 ($\sim 1.3\rm\, mm$), which would require grains of at least $\sim 3\rm\, mm$ in size. 

This discrepancy in the grain sizes estimated from the scattering and ``$\beta$ index" methods needs to be resolved before we can use either method to probe the grain growth
in protoplanetary disks with confidence.
Obviously, one (or both) of the two methods need to be modified to agree with the other. 
In this paper, we will assume that the opacity index $\beta$ is indeed small and that the small $\beta$ comes from big dust grains, and 
focus on understanding the role of such grains in producing 
polarization at shorter wavelengths through scattering. 
In Sec.~\ref{sec:methods}, we present the two methods adopted in this work: semi-analytic method
and Monte Carlo Radiative Transfer. The first method is good to explore a large parameter space quickly, 
whereas the second (more time consuming but more general) method serves as a check on the first and is able to produce polarization maps that can aid 
interpretation of the results. 
In Sec.~\ref{sec:bigdust}, we will first focus on one particular type of dust grains of certain 
compositions and study the behavior of big dust grains under this assumption. 
In Sec.~\ref{sec:phase}, we relax the composition assumption and present the ``phase diagram''
of grain compositions. 
The problems with the ``$\beta$ index'' method are briefly discussed in Sec.~\ref{sec:discussion}, 
together with the implications and limitations of this work. Finally, we give a summary in
Sec.~\ref{sec:summary}.


\section{Method}
\label{sec:methods}
In this work, we will use two methods to study the polarization in a disk. We will first use
a semi-analytical method, called ``Coplanar Isotropic Radiation Field." Under this assumption,
the polarization for a given dust model can be inferred through a simple numerical integral. 
The integral contains the Muller matrix, which will be calculated numerically with Mie
theory \citep{BH83}. We can thus explore the parameter space very quickly. 
We will then use Monte Carlo Radiative Transfer calculations with a simple disk model to 
check our results obtained with the first method. These Monte Carlo Radiative Transfer
calculations also produce polarization maps that can be used to understand the results. 

\subsection{Coplanar Isotropic Radiation Field}
\label{subsec:cirf}
Scattering-induced polarization at (sub)millimeter wavelengths is very different from 
polarization at infrared caused by dust scattering in, e.g., reflection nebulae. The major difference is that the
photons at these wavelengths are dust thermal emission to begin with. When studying the 
polarization from such self-scattering events, one need to, in principle, consider the dust
grains in the whole disk. 
However, even without a global disk model, we can still study the polarization from scattering with
the local \textit{Coplanar Isotropic Radiation Field} (CIRF hereafter) approximation.
This approximation was first introduced in \cite{Yang2016a}, together with Rayleigh scattering approximation
in the small partcile limit\footnote{The CIRF approximation is appropriate near the disk
center and in regions of the disk that are optically thick along the disk plane.}. 
Since it lies at the heart of this work, we will describe the approximation in detail in
the following while relaxing the assumption of small dust grains.

\begin{figure}[ht!]
  \includegraphics[width=0.5\textwidth]{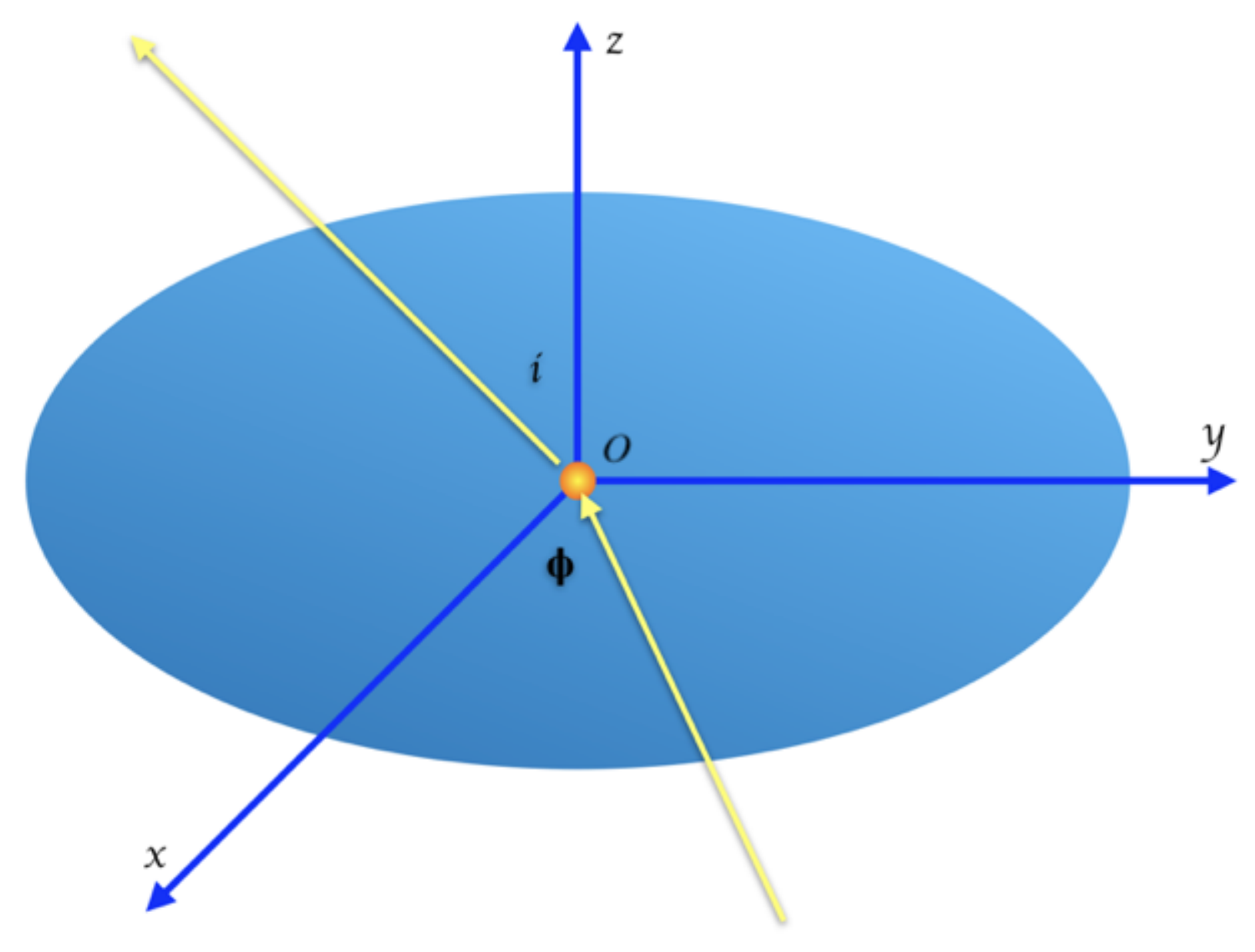}
  \caption{Geometry adopted. A spherical scatterer is placed at the center of the coordinate.
  It is illuminated by unpolarized radiation confined to the $xOy$ plane. The line of sight
  lies in the $xOz$ plane and has an inclination angle $i$ with the $z$ axis.}\label{fig:sketch}
\end{figure}

Let the scatterer be at the origin $O$ of our Cartesian coordinate system (see Fig.~\ref{fig:sketch}). 
We have the disk on the $xOy$ plane. The question is, what is the polarization state
of the light scattered by this particle, when viewed from a line of sight $l$ 
in the $xOz$ plane with an inclination
angle $i$ with respect to the symmetry axis of the disk $z$.
The sky plane is then a plane perpendicular to this direction. 
In general, we need to consider the incoming light from all solid angles. In this
problem of our interest, we can instead consider only those coming from particles sitting
right on the disk midplane, the $xOy$ plane. 
To describe any incoming light, one then need only one argument $\phi$, the angle between
the negative direction of the incoming light and the $x$-axis. 
Note that the $xOy$ plane, or the disk plane, shares one line with the sky plane: the $y$-axis.
If we put a circle in the $xOy$ plane, it will be projected to the sky plane as an ellipse.
This $y$-axis will then be the major axis of this ellipse. A direction perpendicular to
both $y$-axis and the line of sight is then the direction of the minor axis of the ellipse.

We will use Stokes vector $\mathbf{S}\equiv (I,Q,U,V)$ to describe the polarization state of
the light, and further assume all incoming radiation is unpolarized. This is valid as long
as the grains are spherical or not aligned and secondary effects from multiple 
scattering events are negligible. This assumption helps us focus on understanding
the scattering problem, without the complication from grain alignment. 
Under these assumptions, the Stokes $I$ of the scattered light can be expressed as\footnote{This can be made more rigorious with the
aid of Dirac $\delta$ function. In this work, we only care about the ratio
of the Stokes I and Q of the scattered light. So the same factor in the front can be safely
ignored. For examples of more rigorous formulae, see \cite{Yang2016a}.}:
\begin{equation}
  I \propto \int_0^{2\pi} Z_{11}(\theta_s(\phi)) d\phi,
  \label{eq:I}
\end{equation}
where $\theta_s$ is the scattering angle (i.e. the angle between the incoming and scattered
light). For the adopted setting, we have 
$\cos\theta_s(\phi) = -\sin i\cos\phi$.
$Z_{ij}$ is the Muller matrix, with $Z_{11}$ being the component that relates the 
incoming Stokes $I$ with the scattered Stokes $I$. 
Once given the dielectric function and the grain size parameters $x\equiv 2\pi a/\lambda$, 
the Muller matrix can be calculated through the Mie theory \citep{BH83}, assuming compact
spherical dust grains.

The expressions for Stokes $Q$ and $U$ are more complicated:
\begin{equation}
  Q \propto \int_0^{2\pi} Z_{21}(\theta_s(\phi)) \cos(2\phi'(\phi))d\phi,
\end{equation}
\begin{equation}
  U \propto \int_0^{2\pi} Z_{31}(\theta_s(\phi)) \sin(2\phi'(\phi))d\phi,
\end{equation}
where the Stokes $Q$ has been defined such that $Q=I$ implies fully polarized light along
$x'$-direction, the direction of $x$ axis projected to the sky plane (i.e., the minor axis of 
an inclined disk). 
$\phi'$ is the angle $\phi$ (between the $x$-axis and the incoming light, see Fig.~\ref{fig:sketch}) projected 
onto the sky plane.
The extra trigonometric function factors with the argument $2\phi'$ (compared to equation \ref{eq:I}) are due to the translation of Stokes vectors from
the scattering frame to the lab frame. It describes how the scattered light originating from 
different incoming directions contributes to the polarization of the scattered light.
For the adopted geometry, we have
\begin{equation}
\begin{split}
  \cos(\phi'(\phi)) = -\frac{\sin \phi}{\sqrt{\cos^2 i\cos^2\phi+\sin^2\phi}},\\
  \sin(\phi'(\phi)) = \frac{\cos i\cos\phi}{\sqrt{\cos^2 i\cos^2\phi+\sin^2\phi}}.
\end{split}
\end{equation}

We can easily check that $\sin(2\phi'(\phi))=-\sin(2\phi'(-\phi))$. At the same time, 
$\theta_s(\phi)=\theta_s(-\phi)$. So we have $U=0$, regardless of the Muller's matrix.
This is expected due to the symmetry of our setting. 


The polarization fraction of the scattered light is given by:
\begin{equation}\label{eq:ps}
p_s = \frac{\int_0^{2\pi} Z_{21}(\theta_s(\phi)) \cos(2\phi'(\phi))d\phi}{\int_0^{2\pi} Z_{11}(\theta_s(\phi)) d\phi}.
\end{equation}
We have defined the Stokes $Q$ such that a \textit{negative} $p_s$ means the polarization is
along the $y$ axis, i.e. the \textit{major} axis of the disk (see discussions above). A 
\textit{positive} $p_s$ then gives a polarization along the \textit{minor} axis of the 
disk (i.e., the $x'$ direction, which is the $x$-axis projected onto the sky plane).

In the limit of Rayleigh scattering, one recovers the following expression
for scattering-induced polarization (see also Eq.~18 in \citealt{Yang2016a}): 
\begin{equation}
p_s = \frac{\sin^2 i}{2+\sin^2 i}.
\end{equation}
For the rest of this paper, we will consider $45^\circ$ as a representative inclination angle.
At this inclination angle, Rayleigh scattering gives $p_s=0.2$, which means $20\%$ of the 
scattered light is polarized.
Note that this doesn't mean the polarization from self-scattering is as high as $20\%$. One
need to take into account the albedo and the disk structure (see Sec.~\ref{sec:bigdust} for
more detailed discussion) to determine the polarization fraction in a disk. 

\subsection{Fiducial disk model and Monte Carlo Radiative Transfer}
\label{subsec:diskmodel}
In order to verify the results obtained with the semi-analytic CIRF method, 
we also conduct Monte Carlo Radiative Transfer calculations
with publicly available RADMC-3D 
code\footnote{\url{http://www.ita.uni-heidelberg.de/~dullemond/software/radmc-3d/}}. 
To do so, one need a dust model and a disk model (a
density distribution and a temperature distribution). The dust model can be 
described by its dielectric function and a grain size distribution. These will be discussed 
in more detail later. We will first introduce
our fiducial disk model, which applies to all calculations in this paper.

The disk model we adopt is the one used in \cite{Hull2018} for IM Lup. This is one of the best
modeled scattering-induced polarization to date (see comparison between Fig.~4 and Fig.~1-3 in 
\citealt{Hull2018}). This model is a modified version of \cite{Cleeves2016}, which is a viscous disk \citep{L-BP74}
described by:
\begin{equation}
  \Sigma = \Sigma_c \left(\frac{R}{R_c}\right)^{-\gamma}
  \exp\left[-\left(\frac{R}{R_c}\right)^{\alpha}\right],
  \label{eq:Sigma}
\end{equation}
where $\Sigma_c=25\rm\, g/cm^2$ is the column density for gas, and a gas-to-dust ratio of $100$
is adopted. $R_c=100\rm\,AU$ is a characteristic radius. The power-law index for column density
$\gamma=0.3$. In the vertical direction, the disk is set to be in hydrostatic equilibrium with
a midplane temperature prescribed as:
\begin{equation}
  T_m = T_0 \left(\frac{R}{R_c}\right)^{-q},
\end{equation}
where $T_0=70$ K and $q=0.43$. We further reduce the scale height of the dust by a factor of $20$,
since we expect them to be well-settled towards the disk midplane (e.g. the HL Tau disk thickness
has been estimated to be thinner than $1$ AU by \citealt{Pinte2016}). The near-far side 
asymmetry in the polarized intensity predicted for optically and geometrically thick disks are 
generally not observed in late type T Tauri systems with ordered polarization maps, which also
implies geometrically thin disk \citep{Yang2017}.

The MC Radiative Transfer calculations are done on a spherical-polar cooridinate system. The 
radial grid is logarithmically spaced between $5$ AU and 200 AU with 150 cells in between.
The $\theta$ grid is uniformly spaced in a wedge with $64$ cells and a half opening angle
of $0.07$ radian. This gives about $5.8$ scale heights on both sides of the disk at $R_c$.
The $\phi$ grid spans $0$ to $2\pi$ uniformly with $64$ cells.
For each MC run, we use $5\times10^{9}$ photon packages.

Ideally, one would like to calculate the temperature profile self-consistently. 
However, the detailed temperature profile strongly depends on the amount of small dust grains
and the spatial distribution of them, which reprocess the star light. 
At the same time, the observed polarization at (sub)millimeter emission is mostly contributed
by large dust grains near the mid plane of the disk. The small dust grains have limited 
contribution to the observed thermal emission beyond the effects on the temperature.
These two species of dust grains are not directly connected to each other. One has to make
some assumptions to connect them together, or prescribe them separately. For example, one can assume a 
simple power-law model, or assume more complicated distributions such as the steady state 
solution of grain coagulation and fragmentation \citep{Birnstiel2012}. One can also split these
two species completely and have the small dust grains mixed with gas and large dust grains 
settled towards the mid plane (e.g. \citealt{Cleeves2016}). 
Once the dust distribution is prescribed, the temperature
distribution can be calculated with the Monte Carlo method, self-consistently. This process,
however, is more computational expensive and is not much better than simply prescribing a 
temperature distribution since one can change the temperature distribution  by changing the 
distribution of the small grains, which is generally not well constrained observationally. 

On the other hand, the Monte Carlo method discussed in this subsection is only for illustrative
purposes to verify our conclusions from the semi-analytical model and produce spatial distributions 
of the polarization properties, such as orientation and polarization fraction. The expensive Monte 
Carlo calculation of the dust temperature is not necessary for this purpose.



\section{Scattering off big dust grains}
\label{sec:bigdust}

\subsection{Low polarization fraction and potential polarization reversal}
With the two methods established, we are now in a good position to study the polarization produced
by the big dust grains through scattering in an inclined disk.
We will focus on the $870\rm\, \mu m$ wavelength, which is the ALMA Band 7. This is the shortest 
wavelength that ALMA can detect dust polarization at and is
where the clear uniform polarization patterns, which are strong signatures of scattering-induced 
polarization, are observed most frequently \citep{Stephens2017,Hull2018,Dent2019,Bacciotti2018}.
At this wavelength, disk polarization has often been attributed to the self-scattering of 
$\sim100\rm\,\mu m$ dust grains. 
Such grains are too small to produce the small opacity index $\beta$ ($\lesssim 1$) that is 
often inferred at millimeter wavelengths. 

For the dust composition, we adopt the dust mixture used in the DSHARP project \citep{Birnstiel2018}.
It is a mixture of water ice \citep{Warren2008}, refractory organics \citep{Henning1996}, troilite \citep{Henning1996},
and astronomical silicate \citep{Draine2003}. The mass fractions are $0.2$, $0.3966$, $0.0743$, and
$0.3291$, respectively. 
We will focus on one dust composition in this section and vary the grain sizes 
in order to understand the effects of grain sizes. 
In Sec.~\ref{sec:phase}, we will relax this assumption and explore the polarization produced
by dust grains with different compositions.

We will further assume a power-law grain size distribution with a power-law index of $-3.5$ 
(\citealt{MRN77}; MRN-distribution hereafter). A distribution of grains with different sizes,
as oppose to single-sized grains, will help avoiding strong oscillations 
(in both the phase function and in the opacities as functions of grain sizes)
when the size parameter is on the order of unity. Other power-laws will be considered in 
Sec.~\ref{subsec:distribution}.

\begin{figure*}[ht!]
  \includegraphics[width=\textwidth]{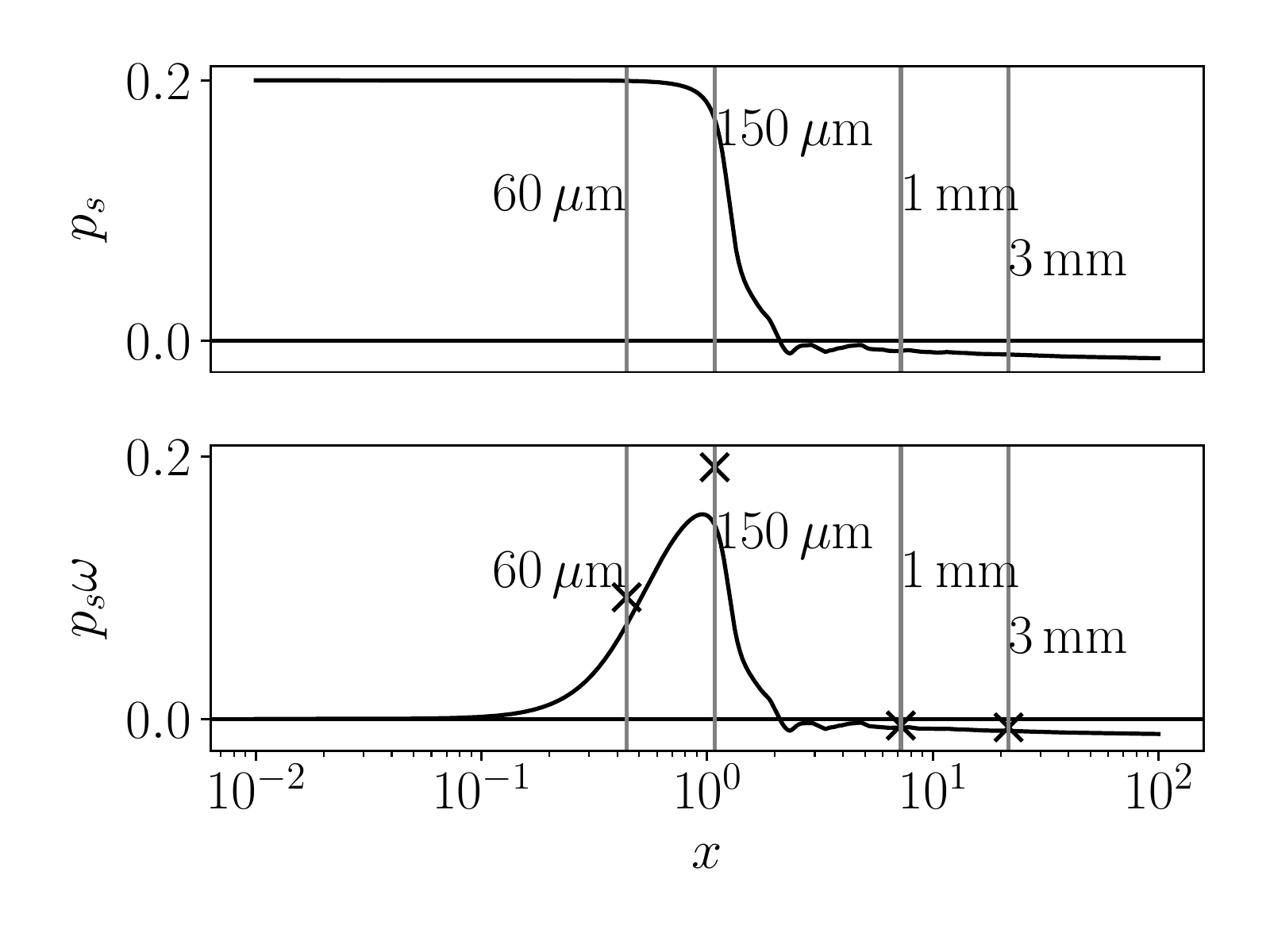}
  \caption{Upper panel: The (signed) polarization fraction of scattered light $p_s$ for a $45^\circ$ inclined disk
  plotted against MRN-distributed dust grains with different maximum size parameters.
  Lower panel: The product of $p_s$ and the albedo $\omega$. 
  Four vertical lines with different size parameters for an observation wavelength of
  $\lambda=870\rm\, \mu m$ are also shown. 
  These are the maximum grain sizes for the four representative models in Sec.~\ref{subsec:pmaps}
  (see Fig.~\ref{fig:pmaps}). The crosses on each of the four lines are the averaged polarization fraction
  in these four models, boosted by the same factor of $10$ for easy comparison with the black solid curve.}
  \label{fig:pvsa}
\end{figure*}

With the dust composition given, we can calculate the Muller Matrix through the Mie theory, 
assuming compact spherical dust grains. We then calculate the polarization from scattering
in an inclined disk numerically through Eq.~\eqref{eq:ps}. 
We will use a fiducial inclination angle $45^\circ$ throughout this paper.
The upper panel of Fig.~\ref{fig:pvsa} shows the (signed polarization fraction) $p_s$ for 
different grain sizes in terms of
the dimensionless size parameter. We can see that the polarization is constant at $20\%$ for 
small dust grains,
but drops very quickly as $x$ increases beyond unity, with some oscillations as it drops. Both the decrease and oscillation are due to different parts of the dust grains
having different phases during the scattering. Interestingly, we find that for big dust grains,
the polarization fraction became negative which, for a disk, means the polarization will be along the \textit{major} axis of the disk. This polarization reversal was
first illustrated in an inclined disk model by \cite{Yang2016a} (see their Fig.~7; see also \citealt{Kataoka2015,Kirchschlager2014}), 
and is the opposite of what is observed.
\cite{Brunngraber2019} also noticed this polarization reversal in the presence of 
big dust grains.

The polarization fraction of the scattered light shown in the upper panel of Fig.~\ref{fig:pvsa}, 
however, is not the polarization fraction
we observe. It will be reduced by the local unpolarized thermal emission, which depends
on the local temperature, the opacity, and the column density. The ratio
of the scattered light to the direct emission doesn't directly depend on the local column density,
because both of them depend on the local column density in the same way. So roughly,
the ratio of the light scattered by a dust grain to the direct thermal emission by that grain depends on the ratio between
the source functions, which is $\kappa_{\rm sca}J_\nu/\kappa_{\rm abs}B_\nu$, with  
$\kappa_\mathrm{sca}$ and $\kappa_\mathrm{abs}$ being the scattering and absorption opacities,
respectively. The $J_\nu$ is the local mean intensity, and the $B_\nu$ is the local black body
radiation intensity.
The observed polarization fraction is then, to the zeroth order, 
$p_s(\kappa_{\rm sca}J_\nu/\kappa_{\rm abs}B_\nu)$. The $J_\nu/B_\nu$ depends on the detailed
disk model, and a radiative transfer calculation is needed to determine its value. 
The $\kappa_{\rm sca}/\kappa_{\rm abs}$ is roughly
the albedo $\omega$ of the dust grain and is solely determined by the dust model. The
lower panel in Fig.~\ref{fig:pvsa} shows the product of $p_s$ and the albedo $\omega$.
This is very similar to Fig.~3 and Fig.~4 in \cite{Kataoka2015}, but our polarization fraction
is averaged over radiation incident on the scatterer from different directions, which is more 
meaningful for an inclined disk.  
\cite{Kataoka2015} used simply the polarization fraction at $90^\circ$ scattering angle,
which is not directly connected to the polarization in an inclined disk.

We can see that small dust grains can hardly produce any polarization due to a small albedo and the corresponding heavy dilution
from direct thermal emission. As we go beyond a size parameter of order of unity, different parts
of the dust grains will have different phases, which causes strong oscillations (see 
\citealt{Tazaki2019a} and their Fig.~7 for an illustration), which in turn cause strong cancellation of 
the polarization, leading to a rapid decrease of the net polarization. 
This creates a peak in the distribution of the product $p_s\omega$ as a function of $x$, 
which was interpreted that the scattering-induced
polarization is sensitive to only dust grains with size parameter on the order of unity. 
For ALMA Band 7, this corresponds to a grain size of about $140\rm\, \mu m$.

For the adopted MRN size distribution with big dust grains, which includes some small dust 
grains with optimal
sizes, the scattering is still dominated by the big ones with little polarization. 
If, for example, the polarization fraction near the peak of the $p_s \omega- x$ curve in the 
lower panel of Fig.~\ref{fig:pvsa} is about $3\sim 4\%$ (based on the numerical results below; see the 
upper-right panel of Fig.~\ref{fig:pmaps}), 
it would be about $0.1\%$ for grains with size parameter of about $5$ or larger. 
As discussed earlier, such large grains are thought to be required to produce the small opacity 
index $\beta$; they would have severe difficulty producing the observed disk polarization at the typical level of $\sim 1\%$.

There are two important assumptions that went into the above calculations: the grain size distribution, and
the dust composition. Big dust grains with different compositions will be studied in Sec.
\ref{sec:phase}. The adopted MRN distribution will be relaxed in Sec.~\ref{subsec:distribution}.
We also assumed compact spherical dust grains. This enables us to calculate optical properties 
of very large dust grains using Mie theory. Irregular or even fluffy dust grains may behave 
differently. Numerical calculations for these dust models, however, are very expansive and
are postponed for future investigation.

\subsection{Monte Carlo calculations and polarization maps}
\label{subsec:pmaps}
\begin{figure*}[ht!]
  \includegraphics[width=\textwidth]{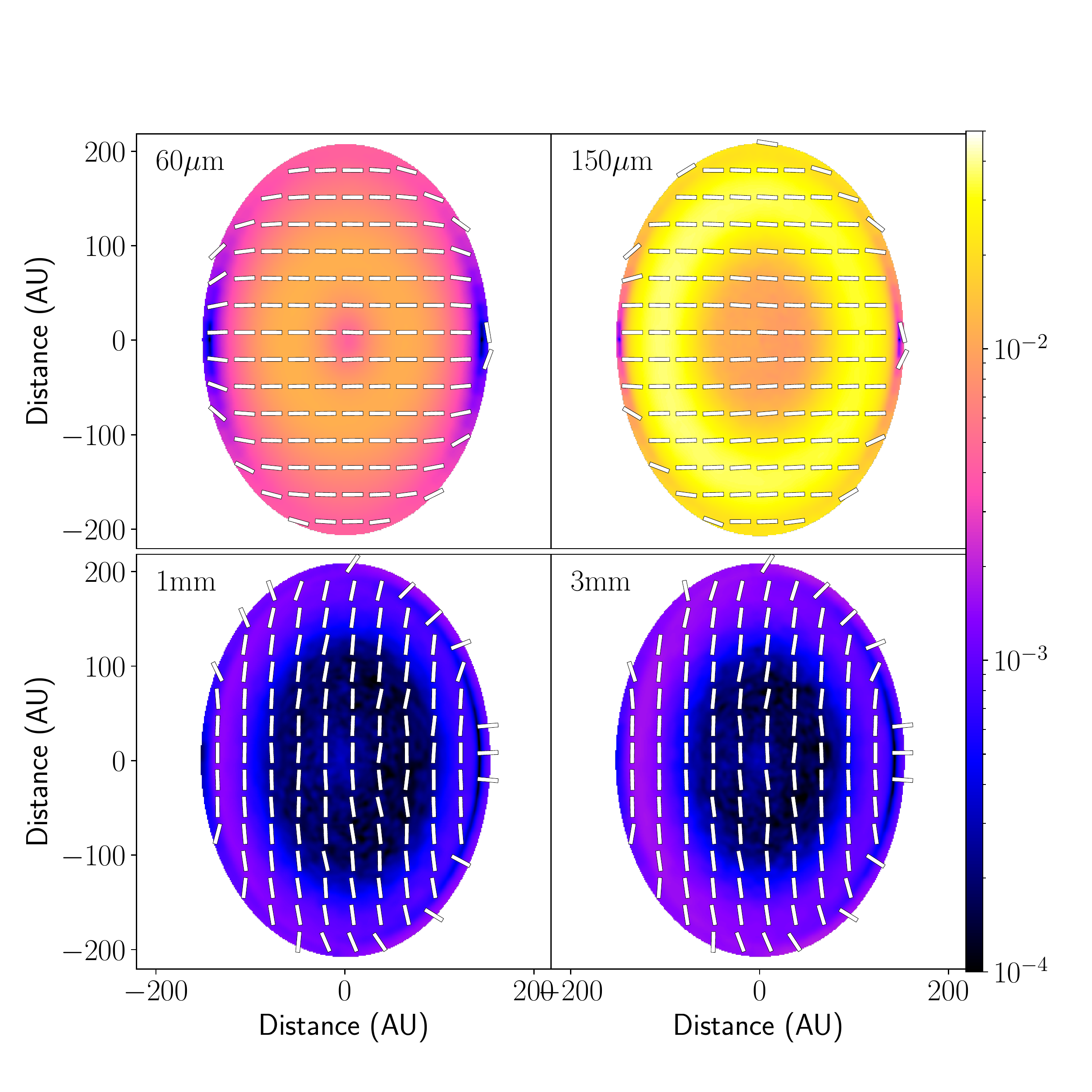}
  \caption{Polarization maps at ALMA Band 7 ($870 \,\rm \mu m$) 
  for the same disk model with different maximum grain sizes: 
  $60 \,\mathrm{\mu m}$, $150 \,\mathrm{\mu m}$, $1$ mm, and $3$ mm. For each panel, the 
  colormap represents the polarization fraction in logrithmic scale. The line segments
  represent the polarization orientation. For the adopted dust model (see text), big 
  dust grains will produce low polarization with polarization reversal.}\label{fig:pmaps}
\end{figure*}

In this subsection, we will verify and extend the semi-analytical results presented in the 
last subsection using Monte Carlo simulations.
For the fiducial disk model described in 
Sec.~\ref{subsec:diskmodel}, we calculate the polarization map with different grain sizes (all with 
MRN distribution, characterized by its maximum grain size $a_\mathrm{max}$), with Monte Carlo 
Radiative Transfer code RADMC-3D. The results are shown in Fig.~\ref{fig:pmaps}. We can see that
the polarization fraction and direction match our semi-analytic results very well: big dust
grains have significantly low polarization fraction. More importantly, the polarization reversal
is clearly seen for the two cases with the largest maximum grain sizes (see the lower two panels 
of Fig.~\ref{fig:pmaps}), although such reversal may not be directly detectable with ALMA, since  
the polarization fraction is only $~0.1\%$.

We also calculated the spectrum index $\alpha$ at wavelengthes between $500 \rm\, \mu m$ and 
$3.1\rm\, mm$\footnote{This range covers the $870\rm\, \mu m$ at which the polarization is calculated.
The calculated spectrum energy distributions are roughly straight lines in logrithmic plots.} 
for the whole disk for different models. Results
are tabulated in Table~\ref{tab:alphas}. We can see that models with $1$ mm maximum grain sizes
or bigger have $\alpha<3$. Under the canonical interpretations, this means $\beta<1$ if it is
optically thin radiation in the Rayleigh-Jeans limit. Our results are therefore consistent with 
the usual interpretation that $\beta < 1$ implies the presence of large grains. 

\begin{table}[ht!]
  \centering
  \begin{tabular}{cccc}
  \hline
    $a_\mathrm{max}$ & $x_\mathrm{max}$ & $\alpha$ & $\beta(\equiv \alpha-2)$\\
  \hline
    60 $\mu$m & 0.43 & 3.18 & 1.18 \\
    150 $\mu$m & 1.08 & 3.29 & 1.29 \\
    1.0 mm & 7.22 & 2.47 & 0.47 \\
    3.0 mm & 21.67 & 2.30 & 0.30 \\
  \hline
  \end{tabular}
  \caption{Spectrum index $\alpha$ (and opacity index $\beta$) for the whole disk with 
  different maximum grain sizes $a_\mathrm{max}$(and the corresponding dimensionless size 
  parameter $x_\mathrm{max}$).}\label{tab:alphas}
\end{table}

For the adopted dust grain composition and disk model, we find that the uniform minor-axis 
polarization and the small spectrum index $\alpha$ are mutually exclusive. Uniform 
minor-axis polarization requires dust grains with size parameter on order of unity or smaller. 
At the same 
time, the inferred small $\beta$ from small spectrum index $\alpha$ corresponds to big dust grains. 
According to \cite{Draine2006}, a size parameter of at least $\sim 15$ is needed to produce a 
$\beta \lesssim 1$. The one order 
difference in grain size estimates from the two methods is a tension that we seek to resolve below.

\section{Polarization phase diagram}
\label{sec:phase}
We have seen that the big dust grain with an oft-adopted composition cannot produce 
enough polarization to explain the observation. Even worse, it may cause polarization reserval, 
which is not firmly detected yet. In this section, we will explore different compositions of dust 
grains, and see if big grains can still produce the observed polarization pattern and fraction.

\subsection{The phase diagram}
For a uniform spherical dust grain, its optical properties can be numerically calculated with the 
Mie theory even if it is very big, e.g. $x \sim 100$, in size. The only input required is the 
complex dielectric function, $\epsilon=\epsilon_r+i\epsilon_i$, or equivalently the 
complex refractive index, $m = n+ik$. 
At the same time, grains with different compositions can be approximated
as a uniform medium with its dielectric function calculated through averaging their ingredients
properly (the so-called ``Effective Medium" method; see \citealt{BH83}). 
As such, we can use the 2D diagram of the complex dielectric function, 
$(n, k)$, to represent any compact spherical dust grain models. 

Now for any point on the phase diagram, we have one type of scatterer represented by
the real and imaginary components of its complex 
refractive index. We will then calculate the optical property of this scatterer assuming MRN 
distribution with maximum size parameter $x_\mathrm{max}=21.67$, which corresponds to a 
maximum grain size of $3\rm\, mm$ for ALMA Band 7.
With the Muller Matrix calculated through the Mie theory \citep{BH83}, 
we can evaluate Eq.~\eqref{eq:ps} 
numerically. This will give the $p_s$, the (signed) polarization fraction for the scattered light 
under the CIRF assumption. From Fig.~\ref{fig:pvsa}, we can see that the peak of the product 
$p_s\omega$ is between $0.15\sim 0.2$. Numerically, one would get about $3\sim 4\%$ polarization
for a $45^\circ$ inclined disk of a representative mass and temperature distribution 
(see the upper right panel of Fig.~\ref{fig:pmaps} for example) using optimal grain sizes.
So in what follows, we will reduce the product $p_s\omega$ by a factor of $5$, to account for the dilution by unpolarized direct thermal emission that is captured in our MC simulations but not by the CIRF method. This MC-calibrated product is more directly comparable to polarization observations.  

\begin{figure*}[ht!]
  \includegraphics[width=\textwidth]{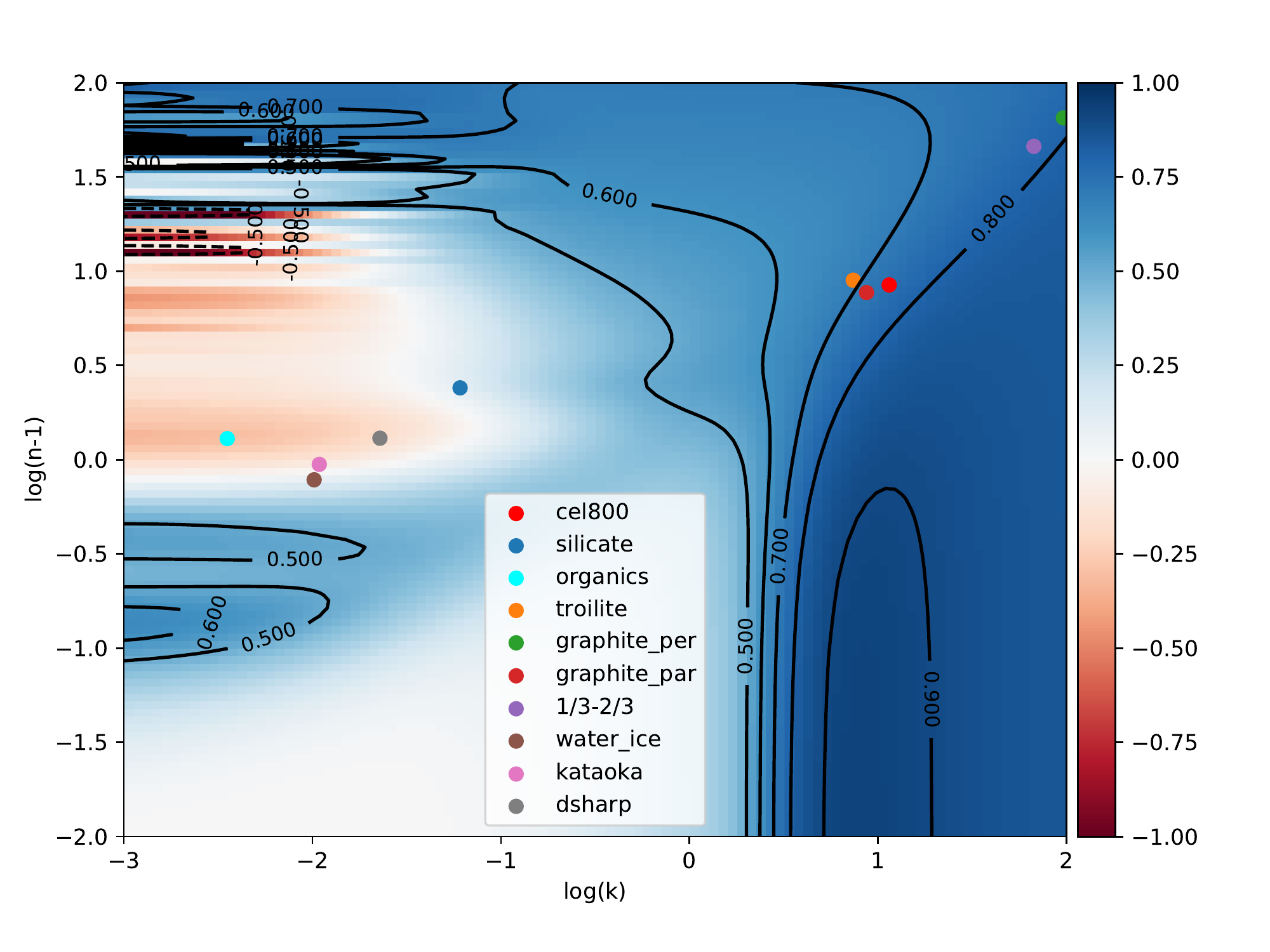}
  \caption{Polarization phase diagram for grains of different optical properties. The dust grains are MRN-distributed with a maximum grain size  of 3 mm and the observing wavelength is ALMA
  Band 7 ($870\rm\, \mu m$). Each point on the map represents compact uniform grains with that effective 
  complex refractive index $m=n+i k$, and the map is presented with $\log_{10}(k)$ and 
  $\log_{10}(n-1)$ as two axes. The color map is polarization percentage (in $\%$). Blue values 
  represent positive polarization fraction, i.e. polarization along minor axis, or the 
  ``normal'' polarization orientation. Red values represent negative polarization fraction,
  i.e.{}\  polarization along major axis, or the ``reversed'' polarization orientation. Several 
  representative dust compositions are marked on the map based on their refractive indexes at
  this wavelength. More detailed description of the dust models is given in 
  Table~\ref{tab:eps_refs}. An inclination angle of $45^\circ$ is assumed. }\label{fig:phase}
\end{figure*}

Fig.~\ref{fig:phase} shows the results for a wide range of complex refractive indices: 
$\log(n-1)$ goes from $-2$ to $2$ and $\log(k)$ from $-3$ to $2$. 
This range covers many types of exotic materials, such as good conductors or strong insulators. 
The color map represents the polarization fraction in a $45^\circ$ 
inclined disk. The blue part has a positive
polarization, meaning that its polarization is along the minor axis of the disk. 
We can see that models with big absorptive dust grains can still produce positive polarization 
with appreciable degree (close to 1 percent). 
There is also region in the parameter space colored in red. Models with dust grains lying in 
these regions produce negative polarization (polarization along major axis), which we call
``polarization reversal".
Contours with selected values of the polarization fraction are also plotted.

To help with the interpretation, we overplot on the phase diagram some dots for the 
refractive indices for different dust grains models with different compositions
at ALMA Band 7 ($870\rm\, \mu m$). 
Their references and brief descriptions are tabulated in Table~\ref{tab:eps_refs}. 
For dust models based on mixture of components, the dielectric functions are averaged using
Bruggleman rule \citep{BH83,Birnstiel2018}.
The \texttt{dsharp} dot
represents the composite dust grains used in Section \ref{sec:bigdust} above. We can see that it lies in the region colored
in red with polarization reversal (see above).
There are four components for this DSHARP mixture: refractory organics, troilite, water ice, and
astronomical silicate. We can see that the refractory organics is the main reason for this 
polarization reversal: it lies deeply in the red region. 
Pure astrophysical silicate doesn't have polarization reversal, but it's polarization fraction is 
small. 
In the upper right corner of the phase diagram lies a couple of dots representing various 
absorptive carbonaceous dust grains. 
In order to produce moderate polarization ($>0.5\%$) using big grains, one need a large fraction of 
such absorptive dust grains. 

\begin{table*}[ht!]
  \begin{center}
  \begin{tabular}{ccccc}
  \hline
  Notation & $n$ & $k$ & Description & Reference \\
  \hline
    \texttt{cel800} & 9.456 & 11.46 & Amorphous carbonaceous dust generated in lab$^*$ & $^a$ \\
    \texttt{silicate} & 3.404 & 0.0607 & Astronomical Silicate & $^b$\\
    \texttt{organics} & 2.293 & 0.00353 & Refractive organics, the so-called ``CHON" particles & $^c$\\
    \texttt{troilite} & 9.954595 & 7.3896 & Troilite & $^c$ \\
    \texttt{graphite\_per} & 66.14 & 96.2 & Graphite for E perpendicular to the c-axis$^\dagger$ &$^b$\\
    \texttt{graphite\_par} & 8.7 & 8.688  & Graphite for E parallel to the c-axis &$^b$\\
    \texttt{1/3-2/3} & 46.99 & 67.03 & Randomly orientated graphite & $^{b,d}$\\
    \texttt{water\_ice} & 1.782 & 0.0102 & Water ice & $^e$\\
    \texttt{kataoka} & 1.944 & 0.0109 & A mixture of dust with organics, silicate and water ice & $^{f,g}$\\
    \texttt{dsharp} & 2.300 & 0.0228 & The DSHARP mixture$^\ddagger$ & $^{h,b,i,j}$\\
  \hline
  \end{tabular}
  \end{center}

  $^*$ See Sec.~\ref{subsec:working} for more detailed description.
  
  $^\dagger$ The c-axis is the axis normal to the ``basal plane" of graphite.

  $^\ddagger$ Dust composition used in the DSHARP collaboration. See Table~1 in \cite{Birnstiel2018} for more details.

  $^a$ \cite{Jager1998}
  $^b$ \cite{Draine2003}
  $^c$ \cite{Pollack1994}
  $^d$ \cite{Draine1993}
  $^e$ \cite{Warren1984}
  $^f$ \cite{Kataoka2015}
  $^g$ \cite{Yang2016a}
  $^h$ \cite{Birnstiel2018}
  $^i$ \cite{Henning1996}
  $^j$ \cite{Warren2008}

  \caption{Representative dust compositions in Fig.~\ref{fig:phase}. The columns list, respectively, the 
  notations used in the legend of Fig.~\ref{fig:phase}, the real and imaginary part of 
  the complex refractive indices at $870\rm\, \mu m$,
  a brief description of the composition, and references.}\label{tab:eps_refs}
\end{table*}

\section{Discussion}
\label{sec:discussion}

\subsection{Different size distributions}
\label{subsec:distribution}
One assumption in our work so far is the MRN dust size distribution. 
It was originally derived for small grains in the interstellar medium \citep{MRN77}. 
As such, it is the most widely adopted assumption for grain size distributions. 
However, there is no direct evidence that the dust grains in protoplanetary disks should follow this
distribution as well. Some authors have suggested a shallower distribution. 
For example, \cite{Birnstiel2012} suggested a complex
size distribution with multiple transition points which is meant to be a steady-state solution to
the grain coagulation problem (see also \citealt{Birnstiel2018}). Near the maximum grain size, this distribution
also suggests a shallower grain size distribution.

\begin{figure}[ht!]
  \includegraphics[width=0.5\textwidth]{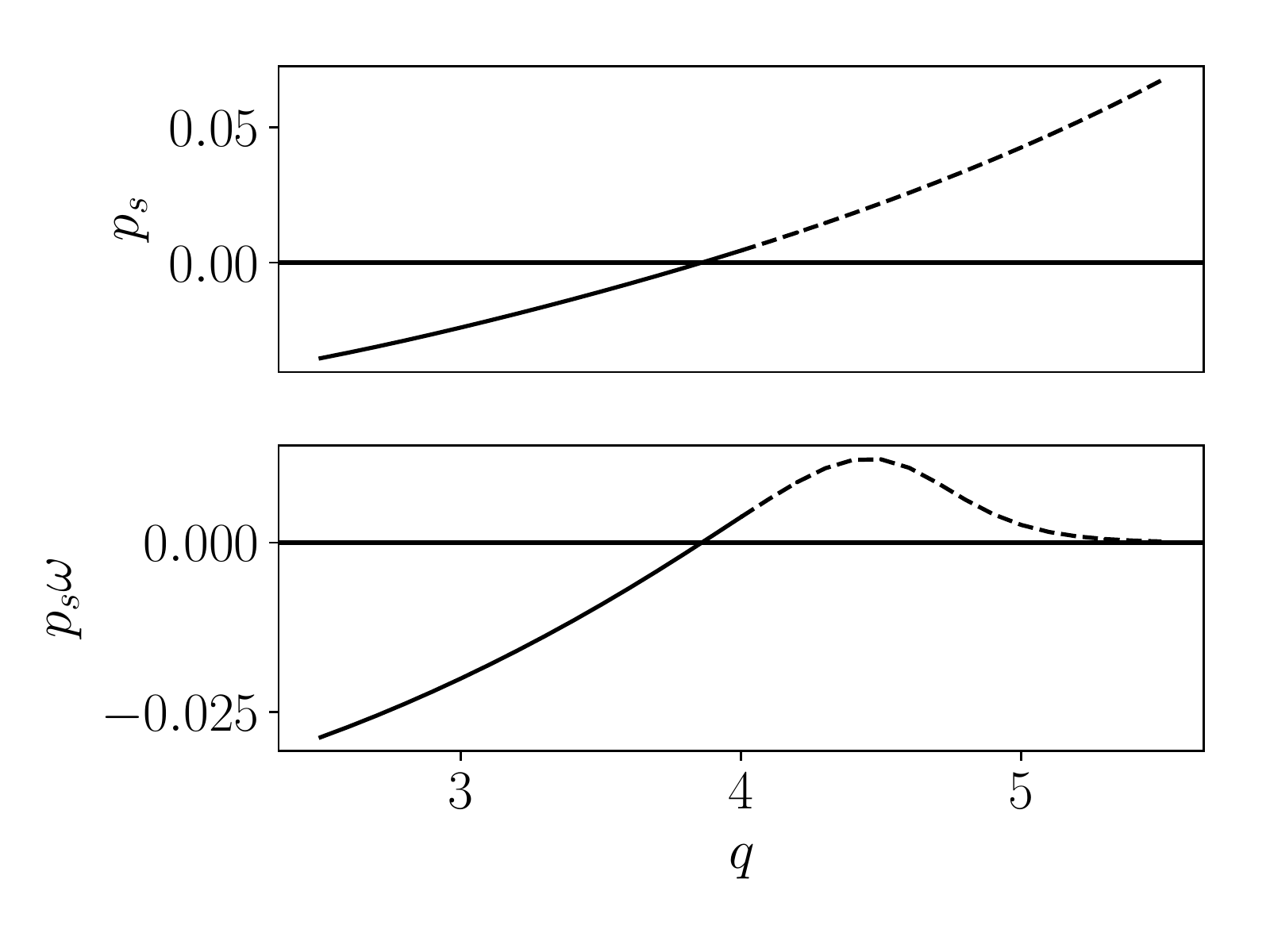}
  \caption{The polarization fraction of scattered light $p_s$ and its product with albedo $\omega$
  as a function of the power-law index $q$. The dashed part represents the part with $q>4$, where the
  most of the mass is in the small dust grains. }\label{fig:diffq}
\end{figure}

In order to study different size distributions, we first replace the MRN assumption with a more
general power-law $n(a)\propto a^{-q}$.
Both the minimum and maximum grain sizes are fixed, with $a_{min}=0.1\rm\, \mu m$ and 
$a_{max}=3$ mm. Fig.~\ref{fig:diffq} shows
the $p_s$ for different power-law indices. 
We can see that as we increase the index $q$, the polarization becomes more positive 
and eventually overcoming the ``polarization reversal'' because of the larger contribution from small grains. The peak of the
observable polarization, indicated by the value of $p_s\omega$, reaches its maximum around $q=4.5$
with an MC-calibrated polarization fraction ($p_s\omega/5$, as in Sec.~\ref{sec:phase}) of only 
$\sim 0.25\%$. We can see that one need to have a much steeper power-law to suppress the polarization reversal and produce a reasonable polarization fraction\footnote{Note that such a steep distribution will have its mass dominated by small particles. In this case, the minimum size parameter may play an important role in the result, but exploring the role in detail is beyond the scope of this paper. }.
%
This is expected because one need to have large fraction of small dust grains ($x\sim 1$) to produce
the desired polarization degree and pattern. A steeper power-law is a natural way to satisfy this requirement.

\subsection{A ``working'' model and polarization at longer wavelengths}
\label{subsec:working}
In Sec.~\ref{sec:phase}, we see that pure absorptive carbonaceous dust grains can potentially produce the desired 
polarization with big grain sizes. In this subsection, we will check this conclusion with our fiducial
disk model and discuss its implications. 

\begin{figure}[ht!]
  \includegraphics[width=0.5\textwidth]{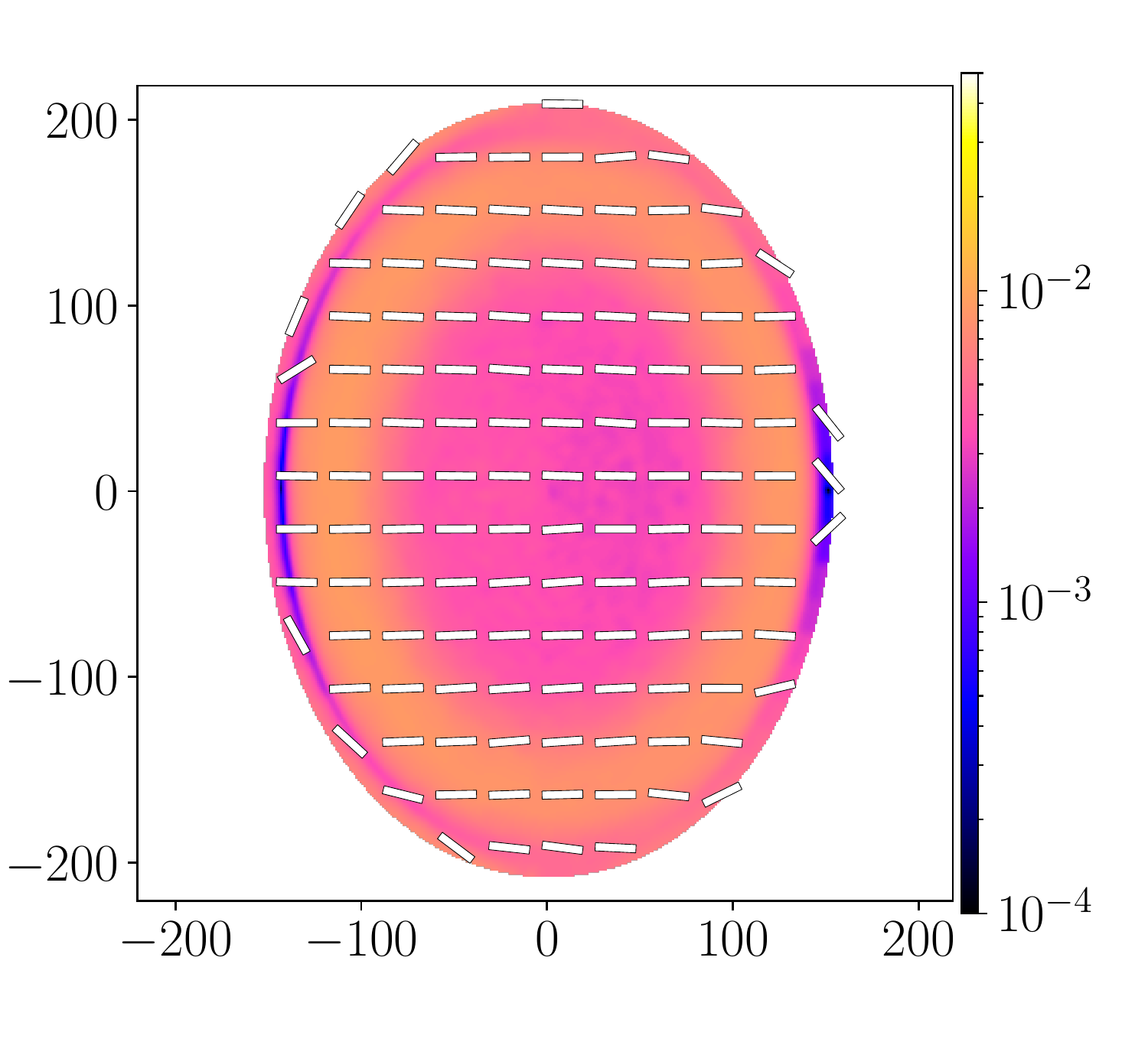}
  \caption{An illustrative model with large pure carbonaceous dust grains. 
  It can produce the correct polarization pattern
  observed at ALMA Band 7, while having a small opacity index $\beta$.}\label{fig:working}
\end{figure}

As an illustrative model, we use the same density and temperature profile described in 
Sec.~\ref{sec:phase}. For the dust grain,  
we use pure amorphous carbonaceous dust grains \citep{Jager1998} with an MRN distribution.
This dust grain is synthesized in a lab through pyrolyzing cellulose materials at 800 $^\circ$C. 
It contains $85.5\%$, $11.9\%$, and $2.6\%$ of Carbon, Hydrogen, and Oxygen atoms, respectively.
The dielectric function was measured and reported up to $\sim 500\rm\, \mu m$ and extrapolated to obtain
dielectric function at (sub)millimeter wavelengths. 
With the maximum grain size set to $3$ mm, we calculate the polarization from this disk at an inclination
angle of $45^\circ$ at the ALMA Band 7 ($870\rm\, \mu m$), as shown in Fig.~\ref{fig:working}.

We can see that the polarization is fairly uniform with no polarization reversal (see Fig.~\ref{fig:working}), unlike the previous models with big dust grains of DSHARP composition (see Fig.~\ref{fig:pmaps}, lower panels). 
The polarization fraction is about $0.6\sim 1\%$ across the disk, which is at the same level
as the one observed in HL Tau Band 7 \citep{Stephens2017}.
The spectrum energy distribution 
was also calculated for the whole disk between $500\rm\, \mu m$ and $3.1$ mm, and the fitted 
SED power-law index $\alpha$ is $2.12$. The inferred opacity index $\beta=0.12$ is
indeed smaller than $1$.
We can see that this model can indeed produce a reasonable polarization degree and pattern while having a small opacity index.

This model is only an illustrative model. Pure carbonaceous dust grains are not very realistic.
In reality, it is possible that certain mixture of dust compositions will have its effective
refractive index lie within the blue region of the phase diagram (Fig.~\ref{fig:phase}). 
It is also possible that big irregular/fluffy dust grains will behave differently than 
what the Mie theory
predicts for compact dust grains. And they may lie effectively in the blue regions, even 
though their complex refractive index doesn't belong there. All these possibilities are 
very interesting to explore and we will postpone them to future investigations.

Another problem of this model is at longer wavelengths. As we see in HL Tau, at the ALMA Band 3
($\sim 3$ mm), the polarization pattern is roughly azimuthal \citep{Kataoka2017}. If one
calculates the polarization for this model at the ALMA Band 3, one would obtain a similar polarization 
pattern with percent level polarization fraction, simply because the grains are large and 
very efficient at producing scattering-induced polarization at the ALMA Band 3.
Since the polarization pattern (with orientations along the minor axis) is not observed, we 
need to address the issue of how this scattering-induced polarization is suppressed at $3$ 
mm. One way is to have bigger grains be better aligned. 
\cite{Draine2009} predicted a transition from no polarization to high polarization going towards
longer wavelengths in dichroic thermal emission of aligned grains based on mixtures of
spheroidal amorphous silicate grains, spheroidal graphite grains, and PAH particles. 
The transition happens around $100\rm\, \mu m$. It is possible to change this transition
wavelength to $\sim 1\rm\, mm$ by changing the grain sizes. Note that their model uses the 
degree of alignment $f$ as a free function of grain sizes, although this is not backed up with detailed alignment theory. Physical justification of the alignment function $f$ at mm/submm grain 
sizes is still needed. On the other hand, even if the big dust grains are perfectly aligned
under some mechanisms, the scattering still cannot be neglected. Calculation of scattering
by aspherical big dust grains and its implementation into radiative transfer calculation
are required to model the Band 3 polarization.

\subsection{Alternatives for resolving the tension in grain size}
In this work, we adopted the conclusion from $\beta$-index measurements that the grain sizes
in protoplanetary disks are big (millimeter in size or bigger).
However, it is possible that the canonical conclusions from $\beta$-index method is not
valid and the dust grains are indeed on the order of $100\rm\, \mu m$. 
The strongest requirement for $\beta$-index method is the medium being optically
thin. This assumption is likely to fail badly in protoplanetary disks at short-wavelength 
ALMA bands, especially
at submillimeter wavelengths. Even though DSHARP survey showed that all their samples are moderately
optically thin, \cite{Zhu2019} pointed out that the conclusion could be wrong and too small by
potentially several orders of magnitude if the albedo of the dust grains is very high. 
The dimming of the disk by scattering has consequences for the observed spectral index
$\alpha$ (and thus the infered opacity index $\beta$ as well), which complicates the picture even further (see also \cite{Liu2019}).

The finding that the grain sizes are about $100\rm\, \mu m$ from all scattering-based polarization 
studies to date appears to require considerable fine-tuning. One would naturally ask: Why are the grain sizes 
always on the order of 
$100\rm\, \mu m$? This question was in part answered by \cite{Okuzumi2019}, who used
the experimental results from \cite{Musiolik2016a,Musiolik2016b} that
the CO$_{2}$ ice mantled dust grains are less sticky and thus yield smaller fragmentation 
barrier to produce a distribution of dust grains with mostly $100\rm\,\mu m$ 
in size in the outer part of the HL Tau disk. The polarization map would fit the observation 
this way as a natural outcome. This tuning in fragmentation barrier changes our 
canonical understanding and need to be investigated in more detail. 

Models with grain sizes capped at $100\rm\, \mu m$ also put some constraints on the 
planet-disk interaction.
The first well-resolved protoplanetary disk image of the HL Tau system from ALMA long baseline 
campaign shows a very 
beautiful disk with rings and gaps \citep{ALMA2015}. At places, the contrast between 
rings and 
gaps could be a factor of a few or even a factor of $10$. If we believe this is thermal 
emission of
$100 \rm\, \mu m$ dust grains, there should be a similar contrast between rings and gaps
for the gas component as well, because $100 \rm\, \mu m$ dust grains are expected to have 
relatively small Stokes numbers\footnote{For a rotating disk, the Stokes number can be expressed as
$\textrm{St}=\rho_s a/\Sigma$, where $\Sigma$ is the column density for gas. For the adopted
disk model prescribed by Eq.~\ref{eq:Sigma}, the Stokes number for $100\rm\, \mu m$ dust grains 
is on the order of
$5\times 10^{-4}$ to $10^{-2}$, depending on the location in the disk.} 
and are thus expected to be well-mixed with gas. 
The depth of the gap opened by a planet depends on various physical properties (see e.g. \citealt{Zhang2018}), 
such as the mass of the planet, the scale height of the disk, the viscosity in the disk.
A factor of 10 depletion of the gas component in the gap would put a strong constraint on the 
allowed parameter space of a planet-disk interaction system.


\section{Summary}
\label{sec:summary}

In this work, we have studied the scattering-induced polarization in inclined disks with grains of different sizes, particular big grains. Our main results are summarized as follows:

\begin{enumerate}
\item We developed a semi-analytical model under the Coplanar Isotropic Radiation Field (CIRF) approximation. It calculates the polarization fraction in scattered light from a particle sitting at the center of an inclined geometrically thin disk.  This model only requires the Muller Matrix of the dust as an input. It does not require detailed radiative transfer and is thus ideal for exploring large parameter space efficiently.

\item With an oft-adopted dust composition, we calculated the polarization fraction of scattered light in an inclined
disk with $45^\circ$ inclination angle with different grain sizes under the CIRF approximation. We found that for large dust 
grains with size parameter $x > 1$, the polarization fraction is very low, and the signed polarization becomes negative, implying that polarization is along the major (rather than minor) axis of the disk, and is thus reversed. There is, however, no clear observational evidence for such polarization reversal in protoplanetary disks to date.

\item For four representative grain sizes, we calculated the polarization maps with Monte Carlo
Radiative Transfer. The polarization fractions and orientations match our expectations from 
the analytical CIRF model well: dust grains with size parameter of one or smaller have the desired 
polarization degree and orientation; models with big dust grains have very low polarization fraction and reversed polarization orientations. We also calculated the spectrum energy index $\alpha$ and 
the corresponding  opacity index $\beta$. In our models, big dust grains are required to produce small $\beta$ index. From these results, we conclude that uniform polarization along minor axis and small $\beta$ index are \textit{mutually exclusive} (assuming canonical dust compositions and compact spherical geometry). 

\item To alleviate the above tension,  we explored a wide range of dust properties, parameterized by the complex refractive index $m=n+ik$ (Fig.~\ref{fig:phase}) under the CIRF approximation, focusing on MRN size distribution with a maximum grain size of $3$ mm and ALMA Band 7 ($870\rm\, \mu m$). 
We find that there is a parameter region that shows polarization 
reversal. The oft-adopted dust models fall in this region and thus fail to explain both the 
observed polarization map and opacity index. We find that the refractory organics
are responsible for such polarization reversal. 

\item On the phase diagram, we find more absorptive dust models (with bigger imaginary part
of dust grains) can produce the desired polarization in both fraction and orientation with big
dust grains. As an illustration, we calculated a model with pure absorptive carbonaceous dust 
grains with maximum grain size of $3$ mm (Fig.~\ref{fig:working}). The polarization pattern  
is uniformly oriented along the minor axis and the polarization fraction is $0.6\sim1.0$\%. 
The inferred opacity index $\beta$ is also small ($0.12$). In this model, the tension between the scattering-induced polarization and the small $\beta$ index is resolved. This example highlights the potential for using polarization to probe not only the sizes but also compositions of dust grains.
\end{enumerate}

\section*{Acknowledgments}

We thank Bruce Draine, Zhaohuan Zhu, Thomas Henning and Christian Eistrup for helpful discussions.
HY acknowledges support by the Institute for Advanced Study. ZYL is supported in part by NASA 80NSSC18K1095 and NSF AST-1716259, 1815784, and 1910106.

\end{document}